\documentclass[12pt]{article}
\usepackage{amssymb}
\usepackage{amsmath}
\makeatletter
\newcommand{\textarrow}[2][1]
  { \settowidth{\@tempdima}{#2}
    \stackrel{#2}
             {\makebox[#1\@tempdima][l]{\rightarrowfill}}
  }
\makeatother

\newlength{\minitwocolumn}
\catcode`\@=11
\long\def\@makefntext#1{
\protect\noindent \hbox to 3.2pt {\hskip-.9pt  
$^{{\eightrm\@thefnmark}}$\hfil}#1\hfill}               %CAN BE USED 

\def\@makefnmark{\hbox to 0pt{$^{\@thefnmark}$\hss}}    %ORIGINAL 
        
\def\ps@myheadings{\let\@mkboth\@gobbletwo
\def\@oddhead{\hbox{}
\rightmark\hfil\eightrm\thepage}   
\def\@oddfoot{}\def\@evenhead{\eightrm\thepage\hfil
\leftmark\hbox{}}\def\@evenfoot{}
\def\sectionmark##1{}\def\subsectionmark##1{}}

\font\eightrm=cmr8

\newcommand{\be}{\begin{equation}}
\newcommand{\ee}{\end{equation}}
\newcommand{\bea}{\begin{eqnarray}}
\newcommand{\eea}{\end{eqnarray}}

\def\bR{\mbox{\boldmath $R$}}

\newcommand{\calC}{{\cal C}}

\newcommand{\calL}{{\cal L}}

\newcommand{\Q}{{\kern.24em\vrule width.04em height1.4ex%
                 depth-.05ex\kern-.26em\mathsf Q}}
\newcommand{\C}{{\kern.24em\vrule width.04em height1.4ex%
                 depth-.05ex\kern-.26em\mathsf C}}

\newcommand{\circD}{\mathring{D}}
\newcommand{\circP}{\mathring{P}}
\newcommand{\circR}{\mathring{R}}
\newcommand{\circpsi}{\mathring{\psi}}

\newcommand{\qgamma}{{\gamma^{\prime}}}
\newcommand{\pgamma}{{\gamma}}

\def\slash#1{\not\!#1}

\begin{document}

%%%%%%%%%%%%%%%%%%%%%%%%%%%%%%%%%%%%%%%%%%%%%%%%%%%%%%%%%%%%%%%%%%
%%%%%%%%%%%%%%%%%%%%%%%% Title %%%%%%%%%%%%%%%%%%%%%%%%%%%%%%%%%%%
%%%%%%%%%%%%%%%%%%%%%%%%%%%%%%%%%%%%%%%%%%%%%%%%%%%%%%%%%%%%%%%%%%

\baselineskip 0.7cm

\begin{titlepage}
%\today
\begin{flushright}
%UT-04-20
\end{flushright}

\vskip 1.35cm
\begin{center}
{\Large \bf
Fermions in (Anti) de Sitter Gravity in Four Dimensions
}
\vskip 1.2cm
Noriaki IKEDA
%${}^1$
\footnote{%${}^1$
E-mail:\ ikeda@yukawa.kyoto-u.ac.jp,
${}^2$ E-mail:\ fukuyama@se.ritsumei.ac.jp} 
and Takeshi FUKUYAMA$^2$%
\vskip 0.4cm
{\it 
${}^1$
%Department of Mathematical Sciences,
College of Science and Engineering,
Ritsumeikan University \\
Kusatsu, Shiga 525-8577, Japan\\
${}^2$
Department of Physics and R-GIRO, Ritsumeikan University,\\ 
Kusatsu, Shiga, 525-8577 Japan
}

\vskip 1.5cm

%\today
%\date{}

\vskip 1.5cm

\begin{abstract}
Fermions in (anti) de Sitter gravity theory in four dimensions are considered. Especially we propose new fermion actions to derive a Weyl or Majorana fermion action if we break the AdS (dS) group to the Lorentz group in curved spacetime.
\end{abstract}
\end{center}
\end{titlepage}

\setcounter{page}{2}

%%%%%%%%%%%%%%%%%%%%%%%%%%%%%%%%%%%%%%%%%%%%%%%%%%%%%%%%%%%%%%%%%%
%%%%%%%%%%%%%%%%%%%%%%%% Article %%%%%%%%%%%%%%%%%%%%%%%%%%%%%%%%%
%%%%%%%%%%%%%%%%%%%%%%%%%%%%%%%%%%%%%%%%%%%%%%%%%%%%%%%%%%%%%%%%%%

\rm

%%%%%%%%%%%%%%%%%%%%%%%%%%%%%%%%%%%%%%%%%%%%%%%%%%%%%%%%%%%%%%%%%%%%%
%%%%%%%%%%%%%%%%%%%%%%%%%%%%%%   SEC  1    %%%%%%%%%%%%%%%%%%%%%%%%%%
%%%%%%%%%%%%%%%%%%%%%%%%%%%%%%%%%%%%%%%%%%%%%%%%%%%%%%%%%%%%%%%%%%%%%
\section{Introduction}
\noindent
(Anti) de Sitter gravity is the gauge theory of gravitation
whose gauge group $G$ is $SO(2,3)$ for anti de Sitter or $SO(1,4)$
for de Sitter
\footnote{We call the 'anti de Sitter group' for $SO(2,3)$ and 
the 'de Sitter group' for $SO(1,4)$.}
\cite{MacDowell:1977jt}\cite{West:1978nd}\cite{Stelle:1979aj}
\cite{Fukuyama:1983hv}.
In this theory, a spacetime metric is not assumed in advance.
We construct a $SO(2,3)$ or $SO(1,4)$ (topological) gauge theory
and breaking the AdS or dS symmetry to the Lorentz group spontaneously 
by a spacetime scaler Higgs field $Z_A$.
This procedure derives a spacetime metric and the Einstein 
gravitational theory with the negative cosmological 
constant for $SO(2,3)$ and
the positive cosmological constant for $SO(1,4)$.
The remaining symmetry $SO(1,3)$ becomes the local Lorentz group 
in the gravitational theory.
This naturally generates a metric and 
derives the gravity 
from gauge theoretical formulation of a field theory without a metric.

%Groups $SO(t,s)$ have different representations
%if natural numbers $t$ and $s$.
%where $t$ and $s$ are natural numbers
Weyl and Majorana spinors appear 
%to construct 
in realistic models.
%since t
The standard model includes chiral matters 
and 
there are interesting predictions that neutrinos
have Majorana nature\cite{Yanagida:1979as}.
However 
the groups $SO(t,s)$'s have different spinor representations 
depending on natural numbers $t$ and $s$.
$SO(2,3)$ and $SO(1,4)$ do not have any Weyl representation.
A Majorana representation is not permitted in $SO(1,4)$.
Moreover 
%terms in 
a fermion quadratic action similar to the Dirac action
in a Majorana spinor generally
contradicts the charge conjugations
in $SO(2,3)$ or $SO(1,4)$
\cite{Kugo:1982bn}\cite{DeAndrade:1994mb}.
%Therefore 
It is nontrivial and 
%difficult 
challenging
to construct Weyl and Majorana spinor actions 
in the AdS (dS) gravity.
% directly.

In this paper, we propose a new mechanism to derive 
a chiral fermion action from the nonchiral action.
Our idea is that we construct a $SO(2,3)$($SO(1,4)$) 
invariant action to derive Weyl or Majorana fermion action when
we break the symmetries to $SO(1,3)$ spontaneously.
%
%we construct a $SO(2,3)$($SO(1,4)$) 
%invariant fermion action 
%and derive a $SO(1,3)$ Weyl or Majorana femion action
%after spontaneous symmetry breaking.

%$SO(1,3)$, $SO(2,3)$ or $SO(1,4)$

The paper is organized as follows. 
In section 2, we review the (anti) de Sitter gravity theory.
In section 3, we explain Dirac fermion actions in the AdS(dS) gravity.
In section 4, we construct $SO(2,3)$ and $SO(1,4)$ actions
which derive Weyl fermion actions in the gravitational theory 
in four dimensions after breaking the symmetries.
In section 5, we do the same process as in section 4 for Majorana
fermions.
%construct a $SO(2,3)$ and $SO(1,4)$ actions which derive
%Majorana fermion actions in the gravitational theory 
%in four dimensions after breaking the symmetries.
Section 6 includes conclusion and discussion.  
In appendix, we summarize formulae of gamma matrices and 
charge conjugations in $SO(1,3)$, $SO(2,3)$ and $SO(1,4)$ groups.
%In appendix B, 

%extra dimensions which are compactified to four dimensions.
%Anomaly free superstring \cite{Witten} with chiral Fermions needs 
%10 dimensional spacetime. 
%This may be looked as top down approach standing on GUT era.
%Apart from superstring, Supersymmetry reqires the extra dimensions 
%for its symmetry breaking mechanism.
%On the other hand, 
%GUT theory consistent with not only low energy observations 
%but also proton decay and gauge coupling unification needs 
%also extra dimensions \cite{Fukuyama2}.
%This is a bottom up approach.
%The latter approach may be divided further to Orbifold GUT 
%(flat spacetime)\cite{Kawamura} 
%and GUT in the AdS$_d$ spacetime (curved spacetime)\cite{Randall}. 
%Among the extra dimensional models, many models assume the 
%universal extra dimension which stats that the matter fields 
%also reside in the extra dimension \cite{Hisano} uiversally.
%Thus we are forced to consider the Fermions in the extra 
%(AdS$_d$) dimensions.
%Fermions in arbitarary Minkowski spacetime was first 
%discussed by \cite{Kugo:1982bn}.

%%%%%%%%%%%%%%%%%%%%%%%%%%%%%%%%%%%%%%%%%%%%%%%%%%%%%%%%%%%%%%%%%%%%%
%%%%%%%%%%%%%%%%%%%%%%%%%%%%%%   SEC  2    %%%%%%%%%%%%%%%%%%%%%%%%%%
%%%%%%%%%%%%%%%%%%%%%%%%%%%%%%%%%%%%%%%%%%%%%%%%%%%%%%%%%%%%%%%%%%%%%
\section{(Anti) de Sitter Gravity}
\noindent
We summarize preliminary results of the (anti) de Sitter gravity
(AdS(dS) gravity).

The (anti) de Sitter gravity is the gauge theory of gravitation
whose gauge group $G$ is $SO(2,3)$ for anti de Sitter or $SO(1,4)$
for de Sitter.
We consider four dimensional spacetime and 
a tangent vector space $\bR^5$ as the internal space.
We define an special internal vector $Z_A= Z_A(x)$ such that
\begin{eqnarray}
Z_1^2 + Z_2^2 + Z_3^2 + Z_4^2 + Z_5^2 = \mp l^2,
\label{definitionofz}
\end{eqnarray}
where the capital Latin letter $A=1,2,3,4,5$ are
internal indices and $x^{\mu}$ are spacetime coordinates.
The signatures in the right hand side in the equation (\ref{definitionofz}) 
are $-l^2$ for $SO(2,3)$ and $+l^2$ for $SO(1,4)$.
$SO(2,3)$($SO(1,4)$) acts to $Z_A$ 
as a symmetry to make the equation (\ref{definitionofz}) invariant.

First we consider the $SO(2,3)$ case.
We consider a connection field $\omega_{\mu AB}$ and define a 
covariant derivative
\begin{eqnarray}
D_{\mu} \psi 
= \left(\partial_{\mu} - \frac{i}{2} \omega_{\mu AB} S_{AB} \right) \psi,
\end{eqnarray}
where
Greek letters are spacetime indices which runs from $1$ to $4$, 
$\mu = 1, 2, 3, 4$,
and $\psi$ is a Dirac fermion.
%Large Roman letters are a $SO(2,3)$ indices which runs $1$ to $5$, 
%for examples $A = 1, 2, 3, 4, 5$.
%Small Roman letters are a $SO(1,3)$ (local Lorentz) 
%indices which runs $1$ to $4$, for examples $a = 1, 2, 3, 4$.
\begin{eqnarray}
S_{AB} = \frac{1}{4i} [\qgamma_A, \qgamma_B],
\end{eqnarray}
is the generator of the $SO(2,3)$ Lie algebra, where
the definition and formulae of gamma matrices $\qgamma_A$ 
appear in the appendix.
The field strength is derived from the equation
\begin{eqnarray}
i [D_{\mu}, D_{\nu}] = - \frac{1}{2} R_{\mu\nu AB} S_{AB}.
\end{eqnarray}
$R_{\mu\nu AB} $ takes the form
\begin{eqnarray}
R_{\mu\nu AB} 
= \partial_{\mu} \omega_{\nu AB}
- \partial_{\nu} \omega_{\mu AB}
- \omega_{\mu AC} \omega_{\nu CB}
+ \omega_{\nu AC} \omega_{\mu CB}.
\end{eqnarray}

We construct a $SO(2,3)$ invariant action 
\begin{eqnarray}
S_{grav} &=& \int d^4x \calL_{grav}
\nonumber \\
&=& \int d^4x
\epsilon^{ABCDE} \epsilon^{\mu\nu\rho\sigma}
\left(\frac{Z_A}{il}\right)
\Biggl[
%\left[
\left(\frac{1}{16 g^2}\right)
R_{\mu\nu BC} R_{\lambda\rho DE} 
\nonumber \\
&& \qquad
+ \sigma(x)
\left\{\left(\frac{Z_F}{il}\right)^2 -1
\right\}
%\left(\frac{Z_A}{il}\right)
D_{\mu} Z_B D_{\nu} Z_C D_{\rho} Z_D D_{\sigma} Z_E
%\right],
\Biggr],
\end{eqnarray}
where 
$\epsilon^{ABCDE}$ and $\epsilon^{\mu\nu\rho\sigma}$ are completely 
antisymmetric tensors with 
$\epsilon^{12345}=1$ and $\epsilon^{1234}=1$, respectively.
It should be remarked that this action is 
topological in the gauge theoretical sense
and that a metric is not introduced ad hoc.
In contrast to the usual gauge theories, this action 
breaks the conformal invariance,
which leads us to the nontrivial dynamics \cite{Fukuyama:2009cm}.
%We note that this action is topological and a metric is not introduced
%ad hoc.

We break the $SO(2,3)$ group to the local Lorentz group 
$SO(1,3)$ as
\begin{eqnarray}
Z_A = (0, 0, 0, 0, il).
\label{breaking1}
\end{eqnarray}
This breaking derives the vierbein $e_{\mu a}$,
\begin{eqnarray}
D_{\mu} Z_A = 
(\partial_{\mu} \delta_{AB} - \omega_{\mu AB}) Z_B
= \left\{
\begin{array}{cc}
- i \omega_{\mu a5} l \equiv e_{\mu a} & \mbox{if} A=a, \\
0 & \mbox{if} A=5, \\
\end{array}
\right.
\end{eqnarray}
where 
the small Latin letters are
$a = 1,2,3,4$.
The field strength is 
\begin{eqnarray}
R_{\mu\nu ab} = \circR_{\mu\nu ab} + \frac{1}{l^2} e_{[\mu a} e_{\nu] b},
\end{eqnarray}
where 
\begin{eqnarray}
\circR_{\mu\nu ab}
= \partial_{\mu} \omega_{\nu ab}
- \partial_{\nu} \omega_{\mu ab}
- \omega_{\mu ac} \omega_{\nu cb}
+ \omega_{\nu ac} \omega_{\mu cb},
\label{fieldstrength}
\end{eqnarray}
is nothing but the gravitational Riemann tensor
and $e_{[\mu a} e_{\nu] b} = e_{\mu a} e_{\nu b} - e_{\nu a} e_{\mu b}$.
$\calL_{grav}$ takes the Einstein gravity form
\begin{eqnarray}
\calL_{grav}
= \partial_{\mu} \calC^{\mu} 
- \frac{e}{16 \pi G} 
\left(\circR + \frac{6}{l^2}\right).
\label{23gravity}
\end{eqnarray}
Here 
$\partial_{\mu} \calC^{\mu}$ is the topological Gauss-Bonnet term.
$e= \det (e_{\mu a})$ and 
$G$ is the gravitational constant derived from 
$16 \pi G = g^2 l^2$.

In the $SO(1,4)$ case, we replace the gamma matrices 
$\qgamma_A$ by $\pgamma_A$.
We consider a connection field 
%Covariant derivative for spinor is
$\omega_{\mu AB}$ and define a 
covariant derivative
\begin{eqnarray}
D_{\mu} \psi 
= \left(\partial_{\mu} - \frac{i}{2} \omega_{\mu AB} s_{AB} \right) \psi,
\end{eqnarray}
where
\begin{eqnarray}
s_{AB} = \frac{1}{4i} [\pgamma_A, \pgamma_B],
\end{eqnarray}
is the generator of the $SO(1,4)$ Lie algebra.

The field strength is derived from the equation
\begin{eqnarray}
i [D_{\mu}, D_{\nu}] = - \frac{1}{2} R_{\mu\nu AB} s_{AB}.
\end{eqnarray}
$R_{\mu\nu AB} $ takes the same form
\begin{eqnarray}
R_{\mu\nu AB} 
= \partial_{\mu} \omega_{\nu AB}
- \partial_{\nu} \omega_{\mu AB}
- \omega_{\mu AC} \omega_{\nu CB}
+ \omega_{\nu AC} \omega_{\mu CB}.
\end{eqnarray}

We construct a $SO(1,4)$ invariant action 
\begin{eqnarray}
S_{grav} &=& - \int d^4x \calL_{grav} 
\nonumber \\
&=& - \int d^4x
\epsilon^{ABCDE} \epsilon^{\mu\nu\rho\sigma}
\left(\frac{Z_A}{l}\right)
\Biggl[
%\left[
\left(\frac{1}{16 g^2}\right)
R_{\mu\nu BC} R_{\lambda\rho DE} 
\nonumber \\
&& \qquad 
+ \sigma(x)
\left\{\left(\frac{Z_F}{l}\right)^2 -1
\right\}
D_{\mu} Z_B D_{\nu} Z_C D_{\rho} Z_D D_{\sigma} Z_E
%\right],
\Biggr],
\end{eqnarray}
We break the $SO(1,4)$ group to the local Lorentz group 
$SO(1,3)$ as
\begin{eqnarray}
Z_A = (0, 0, 0, 0, l).
\label{breaking2}
\end{eqnarray}
This breaking leads to
\begin{eqnarray}
D_{\mu} Z_A = 
(\partial_{\mu} \delta_{AB} - \omega_{\mu AB}) Z_B
= \left\{
\begin{array}{cc}
- \omega_{\mu a5} l \equiv e_{\mu a} & \mbox{if} A=a. \\
0 & \mbox{if} A=5. \\
\end{array}
\right.
\end{eqnarray}
Also the field strength becomes
\begin{eqnarray}
R_{\mu\nu ab} = \circR_{\mu\nu ab} - \frac{1}{l^2} e_{[\mu a} e_{\nu] b},
\end{eqnarray}
where 
$\circR_{\mu\nu ab}$ takes the same form as 
(\ref{fieldstrength}).
%\begin{eqnarray}
%\circR_{\mu\nu ab}
%= \partial_{\mu} \omega_{\nu ab}
%- \partial_{\nu} \omega_{\mu ab}
%- \omega_{\mu ac} \omega_{\nu cb}
%+ \omega_{\nu ac} \omega_{\mu cb},
%\end{eqnarray}
%is the gravitational Riemann tensor.
$\calL_{grav}$ takes the 
%Einstein gravity 
form
\begin{eqnarray}
\calL_{grav}
= \partial_{\mu} \calC^{\mu} 
- \frac{e}{16 \pi G} 
\left(\circR - \frac{6}{l^2}\right).
\label{14gravity}
\end{eqnarray}

The cosmological constant is a negative term
$- \left( + \frac{6}{l^2} \right)$ in the action (\ref{23gravity}) 
for $SO(2,3)$, 
a positive term $- \left( - \frac{6}{l^2} \right) 
= + \frac{6}{l^2}$ in the action (\ref{14gravity}) for
$SO(1,4)$.

%%%%%%%%%%%%%%%%%%%%%%%%%%%%%%%%%%%%%%%%%%%%%%%%%%%%%%%%%%%%%%%%%%%%%
%%%%%%%%%%%%%%%%%%%%%%%%%%%%%%   SEC  3    %%%%%%%%%%%%%%%%%%%%%%%%%%
%%%%%%%%%%%%%%%%%%%%%%%%%%%%%%%%%%%%%%%%%%%%%%%%%%%%%%%%%%%%%%%%%%%%%
\section{Dirac Fermion in AdS(dS) Gravity}
\noindent
In this section, we review the Dirac fermion actions 
in the AdS(dS) gravity theory \cite{Fukuyama:1983hv}.
%This section is the review of the paper 

Let $\psi$ be a $SO(2,3)$($SO(1,4)$) Dirac fermion.
Note that a $SO(2,3)$($SO(1,4)$) Dirac spinor $\psi$ is 
a $SO(1,3)$ Dirac spinor because 
$SO(2,3)$($SO(1,4)$) gamma matrices are the same form as 
$SO(1,3)$ gamma matrices if we add $\gamma_5$.

%%%%%%%%%%%%%%%%%%%%%%%%%%%%%%%%%%%%%%%%%%%%%%%%%%%%%%%%%%%%%%%%%%%%%%%
\subsection{$SO(2,3)$}
\noindent
First we consider the AdS ($SO(2,3)$) gravity.
A $SO(2,3)$ invariant Dirac spinor action is defined 
as \cite{Fukuyama:1983hv}
\begin{eqnarray}
{\cal L}_{Dirac}
= \epsilon^{ABCDE} \epsilon^{\mu\nu\rho\sigma}
{\bar \psi} 
\left(i	S_{AB} \frac{\overleftrightarrow{D}_{\mu}}{3!}
- i \lambda \frac{Z_A}{il} \frac{D_{\mu} Z_B}{4!}
\right) \psi
D_{\nu} Z_C D_{\rho} Z_D D_{\sigma} Z_E, 
\label{dirac1}
\end{eqnarray}
where
${\bar \psi} = \psi^{\dag} \qgamma^5 \qgamma^4$
%${\bar \psi} = \psi^{\dag} \qgamma^4$, 
%\begin{eqnarray}
%D_{\mu} = \partial_{\mu} - \frac{i}{2} \omega_{\mu AB} S_{AB}
%\end{eqnarray}
and
\begin{eqnarray}
\bar{\psi} S_{AB} \overleftrightarrow{D}_{\mu} \psi
= \frac{1}{2} (\bar{\psi} S_{AB} {D}_{\mu} \psi
- \bar{\psi} \overleftarrow{D}_{\mu} S_{AB} \psi).
\end{eqnarray}
Here
\begin{eqnarray}
\bar{\psi} \overleftarrow{D}_{\mu} 
= \bar{\psi}
\left(\overleftarrow{\partial}_{\mu} 
+ \frac{i}{2} \omega_{\mu AB}^* S_{AB} \right).
\end{eqnarray}
We consider always the sum of $L_{grav}$ and fermionic part 
but do not write $L_{grav}$ explicitly hereafter.
%We consider the total Lagrangian ${\cal L}_{grav}+{\cal L}_{Dirac}$.
%We do not write ${\cal L}_{grav}$ explicitly hereafter
but we regard that 
the Lagrangian is always the sum of the gravitational part
${\cal L}_{grav}$ and the fermion part.

%Then,
%\begin{eqnarray}
%{\cal L}_{Dirac}^{\dag}
%= {\cal L}_{Dirac}
%\end{eqnarray}
By the symmetry
breaking (\ref{breaking1}) 
($Z^A=(0,0,0,0,il)$)
from $SO(2,3)$ to $SO(1,3)$,
%\begin{eqnarray
%\label{GF1}
%\end{eqnarray}
${\cal L}_{Dirac}$ reduces to the Dirac action in 
the four dimensional curved spacetime 
\begin{eqnarray}
{\cal L}_{Dirac}
&=& -e {\bar \psi} \left(
\pgamma_a
%e^{\mu a} \overleftrightarrow{\circD}_{\mu}
e^{\mu a} \overleftrightarrow{D}_{\mu}
+ \lambda 
\right) \psi,
= -e {\bar \psi} \left(
\frac{1}{2} 
e^{\mu a} 
\left(
\pgamma_a
\overrightarrow{D}_{\mu}
- \overleftarrow{D}_{\mu}
\pgamma_a
\right) 
+ \lambda 
\right) 
\psi,
\label{dirac131}
\end{eqnarray}
where
%${\bar \psi}^{\prime} = \psi \pgamma_4$ and
$\pgamma_a = i \qgamma_5 \qgamma_a$,
$\pgamma_5 \equiv \qgamma_5$
and
$\bar{\psi} = \psi^{\dag} \pgamma^4$.
%\begin{eqnarray}
%\bar{\psi} S_{AB} \overleftrightarrow{\circD}_{\mu} \psi
%= \frac{1}{2} (\bar{\psi} S_{AB} {\circD}_{\mu} \psi
%- \bar{\psi} \overleftarrow{\circD}_{\mu} S_{AB} \psi)
%\end{eqnarray}
%In our notation,  
(\ref{dirac131}) is hermitian 
${\cal L}_{Dirac}^{\dag}={\cal L}_{Dirac}$.
%and satisfies
%${\cal L}_{Dirac}^T= {\cal L}_{Dirac}^* =- {\cal L}_{Dirac}$.
%
%Then,
%\begin{eqnarray}
%{\cal L}_{Dirac}^{\dag}
%= {\cal L}_{Dirac}
%\end{eqnarray}
The mass of the Dirac spinor is 
%shifted to
\begin{eqnarray}
m = \lambda.
% - \frac{2}{l}.
\end{eqnarray}

%%%%%%%%%%%%%%%%%%%%%%%%%%%%%%%%%%%%%%%%%%%%%%%%%%%%%%%%%%%%%%%%%%%%%%
\subsection{$SO(1,4)$}
\noindent
In the dS $SO(1,4)$ gravity,
we consider a $SO(1,4)$ invariant Dirac spinor action 
\begin{eqnarray}
{\cal L}_{Dirac}
= - \epsilon^{ABCDE} \epsilon^{\mu\nu\rho\sigma}
{\bar \psi} 
\left(\frac{Z_A}{l} \pgamma_{B} \frac{\overleftrightarrow{D}_{\mu}}{3!}
+ \lambda \frac{Z_A}{l} \frac{D_{\mu} Z_B}{4!}
\right) \psi
D_{\nu} Z_C D_{\rho} Z_D D_{\sigma} Z_E,
\label{dirac2}
\end{eqnarray}
which is a slightly different form from the $SO(2,3)$ case.
Here ${\bar \psi} = \psi^{\dag} \pgamma^4$ and
\begin{eqnarray}
\bar{\psi} \pgamma_{B} \overleftrightarrow{D}_{\mu} \psi
= \frac{1}{2} (\bar{\psi} \pgamma_{B} {D}_{\mu} \psi
- \bar{\psi} \overleftarrow{D}_{\mu} \pgamma_{B} \psi).
\end{eqnarray}

%\begin{eqnarray}
%{\cal L}_{Dirac}^{\dag}
%= {\cal L}_{Dirac}
%\end{eqnarray}
By the symmetry
breaking (\ref{breaking2}) 
%$Z^A=(0,0,0,0,l)$
from $SO(1,4)$ to $SO(1,3)$, 
${\cal L}_{Dirac}$ reduces to the Dirac action in 
the four dimensional curved spacetime 
\begin{eqnarray}
{\cal L}_{Dirac}
&=& -e {\bar \psi} \left(
\pgamma_a
%e^{\mu a} \overleftrightarrow{\circD}_{\mu}
e^{\mu a} \overleftrightarrow{D}_{\mu}
+ \lambda 
\right) \psi,
= -e {\bar \psi} \left(
\frac{1}{2} 
e^{\mu a} 
\left(
\pgamma_a
\overrightarrow{D}_{\mu}
- \overleftarrow{D}_{\mu}
\pgamma_a
\right) 
+ \lambda 
\right) 
\psi,
\label{dirac132}
\end{eqnarray}
where $\bar{\psi} = \psi^{\dag} \pgamma^4$.
%where
%${\bar \psi}^{\prime} = \psi \pgamma_4$ 
%and
%$\pgamma_a = i \qgamma_5 \qgamma_a$.
%We can confirm that 
(\ref{dirac132}) is hermitian 
${\cal L}_{Dirac}^{\dag}={\cal L}_{Dirac}$.
%and satisfies
%${\cal L}_{Dirac}^T= {\cal L}_{Dirac}^* =- {\cal L}_{Dirac}$.
In this case, we also have 
a mass
%a complex mass 
\begin{eqnarray}
m = \lambda.
%m = \lambda - \frac{2i}{l}.
\end{eqnarray}

%\vfil\eject
%%%%%%%%%%%%%%%%%%%%%%%%%%%%%%%%%%%%%%%%%%%%%%%%%%%%%%%%%%%%%%%%%%%%%

%%%%%%%%%%%%%%%%%%%%%%%%%%%%%%%%%%%%%%%%%%%%%%%%%%%%%%%%%%%%%%%%%%%%%
%%%%%%%%%%%%%%%%%%%%%%%%%%%%%%   SEC  4    %%%%%%%%%%%%%%%%%%%%%%%%%%
%%%%%%%%%%%%%%%%%%%%%%%%%%%%%%%%%%%%%%%%%%%%%%%%%%%%%%%%%%%%%%%%%%%%%
\section{Weyl Fermion}
\noindent
Since there exists no Weyl spinor in $SO(2n+1)$ representations, 
we cannot construct Weyl spinors in the 
$SO(2,3)$ and $SO(1,4)$ groups.
In order to realize Weyl fermions in the AdS (dS) gravity,
we consider a $SO(2,3)$ ($SO(1,4)$) action to derive 
a Weyl fermion action in the curved spacetime if we break
the $SO(2,3)$ ($SO(1,4)$) symmetry to the $SO(1,3)$ symmetry.

%%%%%%%%%%%%%%%%%%%%%%%%%%%%%%%%%%%%%%%%%%%%%%%%%%%%%%%%%%%%%%%%
\subsection{$SO(2,3)$}
Let $\psi$ be a $SO(2,3)$ Dirac spinor.
%$SO(2,3)
%(SO(1,4))
%$.
We introduce a projection operator,
\bea
P_{\pm}
\equiv \frac{1}{2} \left(1 \pm \frac{Z_A \qgamma^A}{il}\right),
\eea
and define
\begin{eqnarray}
&& \psi_{\pm} \equiv P_{\pm} \psi.
\end{eqnarray}
Since $P_{\pm}$ is $SO(2,3)$ covariant, 
$\psi_{\pm}$ is a covariant spinor.
We can construct a $SO(2,3)$ invariant action 
by modifying (\ref{dirac1}),
\begin{eqnarray}
{\cal L}_{Weyl}
= \epsilon^{ABCDE} \epsilon^{\mu\nu\rho\sigma}
{\bar \psi_+} 
\left(i S_{AB} \frac{\overleftrightarrow{D}_{\mu}}{3!}
- i \lambda \frac{Z_A}{il} \frac{D_{\mu} Z_B}{4!}
\right) \psi_+
D_{\nu} Z_C D_{\rho} Z_D D_{\sigma} Z_E.
\label{weyl1}
\end{eqnarray}
If we break the $SO(2,3)$ symmetry
\begin{eqnarray}
Z^A=(0,0,0,0,il),
%\label{GF1}
\end{eqnarray}
$P_{\pm}$ reduce to the chiral projections $\circP_{\pm}$
\begin{eqnarray}
P_{\pm} \longrightarrow \circP_{\pm} = \frac{1 \pm \qgamma_5}{2}
= \frac{1 \pm \pgamma_5}{2}.
\end{eqnarray}
Then $\psi_+$ becomes Weyl spinors $\circpsi_{\pm}$
\begin{eqnarray}
&& \psi_{\pm} \longrightarrow \circpsi_{\pm} = \circP_{\pm} \psi,
\end{eqnarray}
which have definite chirality respectively.
%according to signs.

The action (\ref{weyl1}) becomes
a $SO(1,3)$ massless Weyl fermion action
\begin{eqnarray}
{\cal L}_{Weyl}
&=& -e {\bar \circpsi_+} \left(
\pgamma_a
e^{\mu a} \overleftrightarrow{D}_{\mu}
+ \lambda 
\right) \circpsi_+
%\nonumber \\ &=&
= -e {\bar \circpsi_+} \left(
\pgamma_a
e^{\mu a} \overleftrightarrow{\circD}_{\mu}
\right) \circpsi_+,
%\label{Weyl2}
\end{eqnarray}
where
\begin{eqnarray}
\circD_{\mu} = \partial_{\mu} - \frac{i}{2} \omega_{\mu ab} S_{ab}
= \partial_{\mu} - \frac{i}{2} \omega_{\mu ab} s_{ab}.
%\label{diracreduction}
\end{eqnarray}
Since the chiral projections make the two components to zero
among four components of Dirac spinors, the mass term 
automatically drops out.

%%%%%%%%%%%%%%%%%%%%%%%%%%%%%%%%%%%%%%%%%%%%%%%%%%%%%%%%%%%%%%%%%%%%%%%%%
\subsection{$SO(1,4)$}
\noindent
Let $\psi$ be a $SO(1,4)$ Dirac spinor.
In the $SO(1,4)$ case, we introduce
\bea
P_{\pm}
\equiv \frac{1}{2} \left(1 \pm \frac{Z_A \pgamma^A}{l}\right),
\eea
and define
\begin{eqnarray}
&& \psi_{\pm} \equiv P_{\pm} \psi.
\end{eqnarray}
Since $P_{\pm}$ is $SO(1,4)$ covariant,
$\psi_{\pm}$ is a covariant spinor.
We can construct $SO(1,4)$ invariant action 
by modifying 
(\ref{dirac2}),
\begin{eqnarray}
{\cal L}_{Weyl}
= - \epsilon^{ABCDE} \epsilon^{\mu\nu\rho\sigma}
{\bar \psi_+} 
\left(\frac{Z_A}{l} \pgamma_{B} \frac{\overleftrightarrow{D}_{\mu}}{3!}
+ \lambda \frac{Z_A}{l} \frac{D_{\mu} Z_B}{4!}
\right) \psi_+
D_{\nu} Z_C D_{\rho} Z_D D_{\sigma} Z_E.
\label{weyl3}
\end{eqnarray}
If we break the $SO(1,4)$ symmetry as
\begin{eqnarray}
Z^A=(0,0,0,0,l),
\end{eqnarray}
$P_{\pm}$ reduce to chiral projections $\circP_{\pm}$
\begin{eqnarray}
P_{\pm} \longrightarrow \circP_{\pm} = \frac{1 \pm \pgamma_5}{2}.
\end{eqnarray}
Then $\psi_+$ becomes Weyl fermions $\circpsi_{\pm}$,
\begin{eqnarray}
&& \psi_{\pm} \longrightarrow \circpsi_{\pm} = \circP_{\pm} \psi,
\end{eqnarray}
which have definite chirality respectively.
%according to signs.

The action (\ref{weyl3}) becomes
a $SO(1,3)$ massless Weyl fermion action
\begin{eqnarray}
{\cal L}_{Weyl}
&=& -e {\bar \circpsi_+} \left(
\pgamma_a
e^{\mu a} \overleftrightarrow{D}_{\mu}
+ \lambda 
\right) \circpsi_+
%\nonumber \\ &=&
= -e {\bar \circpsi_+} \left(
\pgamma_a
e^{\mu a} \overleftrightarrow{\circD}_{\mu}
\right) \circpsi_+.
%\label{Weyl2}
\end{eqnarray}
Again the mass term drops out.

%\vfil\eject
%%%%%%%%%%%%%%%%%%%%%%%%%%%%%%%%%%%%%%%%%%%%%%%%%%%%%%%%%%%%%%%%%%%%%

%%%%%%%%%%%%%%%%%%%%%%%%%%%%%%%%%%%%%%%%%%%%%%%%%%%%%%%%%%%%%%%%%%%%%
%%%%%%%%%%%%%%%%%%%%%%%%%%%%%%   SEC  5    %%%%%%%%%%%%%%%%%%%%%%%%%%
%%%%%%%%%%%%%%%%%%%%%%%%%%%%%%%%%%%%%%%%%%%%%%%%%%%%%%%%%%%%%%%%%%%%%
\section{Majorana Fermion}
\noindent
A Majorana fermion $\psi_M$ in four dimensional spacetime
with the local Lorentz symmetry is defined by
\begin{eqnarray}
\psi_M = \psi_M^c \equiv C \bar{\psi}_M^T,
%= \qgamma_2 \psi_M^*
\label{majorana0}
\end{eqnarray}
where $C$ is the charge conjugation in four dimensional spacetime.
If we take the Dirac (Pauli) basis, $C$ is 
\begin{eqnarray}
C = \pgamma_2 \pgamma_4.
\label{chargeconj1}
\end{eqnarray}
%First we consider $SO(1,3)$ gamma matrices $\pgamma_a$ 
%In this basis, the $SO(1,3)$ charge conjugation $C$ is defined as 
However $C$ is not covariant under either $SO(2,3)$ or $SO(1,4)$.
$\psi_M$ is not consistent with the $SO(2,3)$ ($SO(1,4)$) covariance.

%In order to define a Majorana spinor, we 
%define the 'charge conjugation' $C$.
In the following, we consider the consistent $SO(2,3)$ ($SO(1,4)$) 
'charge conjugation' and an
action to derive a Majorana fermion action by breaking
the $SO(2,3)$ 
($SO(1,4)$) 
symmetry to the $SO(1,3)$ symmetry.

%%%%%%%%%%%%%%%%%%%%%%%%%%%%%%%%%%%%%%%%%%%%%%%%%%%%%%%%%%%%%%%%%%%
\subsection{$SO(2,3)$}
\noindent
The $SO(2,3)$ gamma matrices $\qgamma_A$ 
are constructed as
\begin{eqnarray}
&& 
\qgamma_a \equiv - i \pgamma_5 \pgamma_a,
\nonumber \\
&& 
\qgamma_5 \equiv \pgamma_5,
\end{eqnarray}
which appear the action (\ref{dirac1})
of the $SO(2,3)$ Dirac spinor in section 3.
%from construction of the Dirac spinor action in section 3.

%In order to construct a Majorana spinor,
The $SO(2,3)$ charge conjugation $C$ has to preserve 
$SO(2,3)$ covariance of spinors
of $\psi_M$ and $C \bar{\psi}_M^T$,
From this requirement and properties 
of $SO(2,3)$ gamma matrices $\qgamma_A$
in the appendix,
we have two candidates
\footnote{
If the Dirac basis is taken,
$C_1 = C_2$
since the formulae (\ref{gamma5}) is satisfied.
However (\ref{gamma5}) is not assumed 
in the definition of $SO(2,3)$ group, 
We do not use (\ref{gamma5}) 
in order to define a charge conjugation under lesser assumptions.
The same discussion is also made in the $SO(1,4)$ case.
}
\begin{eqnarray}
&& C_1 = \qgamma_1 \qgamma_3 \qgamma_5,
\nonumber \\
&& C_2 = \qgamma_2 \qgamma_4.
\end{eqnarray}
Here $C_2 = \qgamma_2 \qgamma_4 = \pgamma_2 \pgamma_4$ is 
equal to the $SO(1,3)$ charge conjugation (\ref{chargeconj1}).
Therefore we can take $C = C_2$ as the $SO(2,3)$ charge conjugation
\footnote{Note that $C = C_2$ is not the same 
as the charge conjugation in the  $SO(2,3)$ spacetime symmetry 
in five dimensions.}.
$C_2$ satisfies 
\begin{eqnarray} 
&& C_2^T = -C_2,
%\nonumber \\ && 
\qquad 
C_2^* = C_2,
%\nonumber \\ && 
\qquad 
C_2^2 = -1,
%\nonumber \\ && 
\qquad 
C_2^{-1} = -C_2,
%\end{eqnarray}
%\begin{eqnarray}
\nonumber \\ && 
C_2^{-1} \qgamma_{A} C_2 = \qgamma_{A}^{T},
%\end{eqnarray}
%\begin{eqnarray}
\qquad C_2^{-1} S_{AB} C_2 = - S_{AB}^T.
\label{c2condition}
\end{eqnarray}
AdS 'Majorana' fermion $\psi_M$ is defined by
\begin{eqnarray}
\psi_M = C \bar{\psi}_M^T = C_2 \bar{\psi}_M^T.
%= \qgamma_2 \psi_M^*
\label{majorana1}
\end{eqnarray}

We propose a $SO(2,3)$ invariant AdS 'Majorana' fermion action 
by replacing a Dirac spinor to an AdS 'Majorana' spinor
in the action (\ref{dirac1})
\begin{eqnarray}
{\cal L}_{Majorana}
= \epsilon^{ABCDE} \epsilon^{\mu\nu\rho\sigma}
{\bar \psi_M} 
\left(i 
S_{AB} \frac{\overleftrightarrow{D}_{\mu}}{3!}
- i \lambda \frac{Z_A}{il} \frac{D_{\mu} Z_B}{4!} 
\right) \psi_M
D_{\nu} Z_C D_{\rho} Z_D D_{\sigma} Z_E.
\label{majoranaaction1}
\end{eqnarray}
Let us investigate consistency of this action.
Substituting (\ref{majorana1}) to the right hand of
(\ref{majoranaaction1}),
we obtain
\begin{eqnarray}
%{\cal L}_{Majorana}
%= 
\epsilon^{ABCDE} \epsilon^{\mu\nu\rho\sigma}
\left(\psi_M^T (C^{T})^{-1} \right)
\left(i 
S_{AB} \frac{\overleftrightarrow{D}_{\mu}}{3!}
- i \lambda \frac{Z_A}{il} \frac{D_{\mu} Z_B}{4!} 
\right) \left(C \bar{\psi}_M^T \right)
D_{\nu} Z_C D_{\rho} Z_D D_{\sigma} Z_E.
\label{majoranatrans1}
\end{eqnarray}
%Transposing the equation (\ref{majoranatrans1}) and make
%integration by part:
We can easily check that 
(\ref{majoranatrans1}) is equal to 
(\ref{majoranaaction1}) as follows:
\begin{eqnarray}
%{\cal L}_{Majorana}
%&=& 
&& 
= - \epsilon^{ABCDE} \epsilon^{\mu\nu\rho\sigma}
{\bar \psi_M} 
\left(- i C^T 
\left(S_{AB} \frac{\overleftrightarrow{D}_{\mu}}{3!} \right)^T
C^{-1} 
- i \lambda C^T 
C^{-1}
\frac{Z_A}{il} \frac{D_{\mu} Z_B}{4!}
\right) \psi_M
%\nonumber \\
%&& 
%\times 
D_{\nu} Z_C D_{\rho} Z_D D_{\sigma} Z_E
\nonumber \\
&& 
= \epsilon^{ABCDE} \epsilon^{\mu\nu\rho\sigma}
{\bar \psi_M} 
\left(i 
S_{AB} \frac{\overleftrightarrow{D}_{\mu}}{3!}
- i \lambda \frac{Z_A}{il} \frac{D_{\mu} Z_B}{4!} 
\right) \psi_M
D_{\nu} Z_C D_{\rho} Z_D D_{\sigma} Z_E
%\nonumber \\
%&& 
= {\cal L}_{Majorana}.
\end{eqnarray}
%where 
%\begin{eqnarray}
%D^T_{\mu} = \partial_{\mu} - \frac{i}{2} \omega_{\mu AB} S_{AB}^T.
%\end{eqnarray}
%If we use 
%the action becomes
Here we have used the Grassmann property of $\psi_M$, 
the identities of $C_2$ and the gamma matrices 
(\ref{c2condition}).
Thus 
(\ref{majoranatrans1}) is equal to 
(\ref{majoranaaction1}) and 
%the definition of the charge conjugation is consistent with the action.
%\begin{eqnarray}
%{\cal L}_{Majorana}^T
%&=& \epsilon^{ABCDE} \epsilon^{\mu\nu\rho\sigma}
%{\bar \psi_M} 
%\left(i S_{AB} \frac{\overleftrightarrow{D}_{\mu}}{3!}
%+ i \lambda \frac{Z_A}{l} \frac{D_{\mu} Z_B}{4!}
%\right) \psi_M
%D_{\nu} Z_C D_{\rho} Z_D D_{\sigma} Z_E.
%\end{eqnarray}
%Therefore,
%$
%{\cal L}_{Majorana}^T 
%= {\cal L}_{Majorana}.
%$
%Since ${\cal L}_{Majorana}^T$ and ${\cal L}_{Majorana}$ lead to the same 
%equations of motion,
the definition of the charge conjugation is consistent with the action
\footnote{In our notation,
a $SO(1,3)$ Majorana fermion action in flat Minkowski spacetime in 
four dimensions is
\begin{eqnarray}
{\cal L}_{Majorana}
= - {\bar \psi_M} ( \pgamma^{\mu} \partial_{\mu} + \lambda )\psi_M,
\end{eqnarray}
%satisfies
%$
%{\cal L}_{Majorana}^T 
%= {\cal L}_{Majorana},
%$
where $C=\pgamma_2 \pgamma_4$ is the $SO(1,3)$ charge conjugation.}.

%We must take $\lambda =0$.

If we break the $SO(2,3)$ symmetry by $Z_A=(0,0,0,0,il)$,
(\ref{majoranaaction1}) reduces to
a $SO(1,3)$ Majorana fermion action 
in the Einstein gravitational theory
in four dimensions
\begin{eqnarray}
{\cal L}_{Majorana}
&=& -e {\bar \psi_M} \left(
\pgamma_a
e^{\mu a} \overleftrightarrow{D}_{\mu}
+ \lambda \right) \psi_M.
%\nonumber \\
%&=& -e {\bar \psi_M} \left(
%\pgamma_a
%e^{\mu a} \overleftrightarrow{\circD}_{\mu} + \lambda
%- \frac{2}{l} \right) \psi_M
\end{eqnarray}
%
%\begin{eqnarray}
%{\cal L}_{Majorana}^T = - {\cal L}_{Majorana}
%\end{eqnarray}
%
The Majorana mass is 
\begin{eqnarray}
m = \lambda.
% - \frac{2}{l}.
\end{eqnarray}

%%%%%%%%%%%%%%%%%%%%%%%%%%%%%%%%%%%%%%%%%%%%%%%%%%%%%%%%%%%%%%%%%%%%%%%
\subsection{$SO(1,4)$}
\noindent
dS 'Majorana' fermion $\psi_M$ is also defined by 
\begin{eqnarray}
\psi_M = C \bar{\psi}_M^T,
%= \qgamma_2 \psi_M^*
\label{majorana2}
\end{eqnarray}
where $C$ is a $SO(1,4)$ 'charge conjugation'.
Let us take two candidates for the 'charge conjugation' 
%are taken
from the $SO(1,4)$ covariance of $\psi_M$ and $C \bar{\psi}_M^T$,
\begin{eqnarray}
&& C_3 \equiv \pgamma_1 \pgamma_3,
\nonumber \\
&& C_4 \equiv \pgamma_2 \pgamma_4 \pgamma_5.
%\nonumber \\
%&& C_5 \equiv \frac{Z_A \pgamma^A}{l} \pgamma_2 \pgamma_4 \pgamma_5,
\end{eqnarray}
$B$ defined by
%for a charge conjugation $C$ as
\begin{eqnarray}
B \psi_M^* = C \bar{\psi}_M^T,
\end{eqnarray}
must satisfy $B^* B=1$ since a charge conjugation 
has a $Z_2$ symmetry.
If we define $B_3$ as
\begin{eqnarray}
B_3 \psi^* = C_3 \bar{\psi}^T,
\end{eqnarray}
\begin{eqnarray}
&& B_3 = \pgamma_1 \pgamma_3 \pgamma_4,
%\nonumber \\ && 
\qquad B_3^* = \pgamma_1 \pgamma_3 \pgamma_4,
\nonumber \\ && 
%\end{eqnarray}
%\begin{eqnarray}
B_3^* B_3 = -1.
\label{b3relation}
\end{eqnarray}
$B_4$ defined by 
\begin{eqnarray}
B_4 \psi^* = C_4 \bar{\psi}^T,
\end{eqnarray}
satisfies 
\begin{eqnarray}
&& B_4 = - \pgamma_2 \pgamma_5, 
%\nonumber \\ && 
\qquad 
B_4^* = - \pgamma_2 \pgamma_5,
\nonumber \\ && 
B_4^* B_4 = -1.
\end{eqnarray}
%We find 
%However 
Since $B$'s constructed from both $C_3$ and $C_4$ satisfy
$B^* B=-1$,
neither $C_3$ nor $C_4$ can be defined
as a consistent charge conjugation.
We could also consider more general candidates 
covariant to $SO(1,4)$:
\begin{eqnarray}
&& C_3^{\prime} \equiv e^{i \theta_3} \pgamma_1 \pgamma_3,
\nonumber \\
&& C_4^{\prime} \equiv  e^{i \theta_4} \pgamma_2 \pgamma_4 \pgamma_5,
\end{eqnarray}
where $\theta_3$ and $\theta_4$ are phase factors.
However $B$'s constructed from both $C_3^{\prime}$ and $C_4^{\prime}$ 
also satisfy $B^* B=-1$ and
we cannot define a consistent charge conjugation from 
$C_3^{\prime}$ or $C_4^{\prime}$.

Now we consider a third candidate:
\begin{eqnarray}
C_5 \equiv \frac{Z_A \pgamma^A}{l} \pgamma_2 \pgamma_4 \pgamma_5.
\end{eqnarray}
$C_5$ satisfies
\begin{eqnarray}
&& C_5^T = -C_5,
%\nonumber \\ && 
\qquad C_5^* = \pgamma_5 \pgamma_4 \pgamma_2 \frac{Z_A^* \pgamma^A}{l},
%\nonumber \\ && C_5^2 = 
%\nonumber \\ && 
\qquad C_5^{-1} = 
\pgamma_5 \pgamma_4 \pgamma_2 \frac{Z_A \pgamma^A}{l},
\nonumber \\ && 
%\end{eqnarray}
%\begin{eqnarray}
C_5 \pgamma_{A}^{T} C_5^{-1}
= 2 \frac{Z_B \pgamma_B}{l} \frac{Z_A}{l}
- \pgamma_{A},
\label{c5condition}
\end{eqnarray}
$\psi_M$ and $C_5 \bar{\psi}_M^T$ are covariant under 
the $SO(1,4)$ group
and if we define 
\begin{eqnarray}
B_5 \psi_M^* = C_5 \bar{\psi}_M^T,
\end{eqnarray}
%from $C= C_5$,
$B_5 =- \frac{Z_A \pgamma^A}{l} \pgamma_2 \pgamma_5$
and 
$B_5^* = \pgamma_2 \pgamma_5 \frac{Z_A \pgamma^A}{l}$
%we can confirm that 
satisfy the $Z_2$ symmetry condition $B_5^* B_5=1$.
Moreover if we break the $SO(1,4)$ symmetry to $SO(1,3)$ as
(\ref{breaking2}), 
we obtain the $SO(1,3)$ charge conjugation
\footnote{
Note that $C = C_5$ is not
%can not be considered as
a charge conjugation in the $SO(1,4)$ spacetime symmetry 
in five dimensions.
%a 'charge conjugation' in the dS gravity theory in four dimensions but 
}
as
\begin{eqnarray}
C_5 \longrightarrow \pgamma_2 \pgamma_4 =C.
%\label{majorana2}
\end{eqnarray}
Therefore the $SO(1,4)$ charge conjugation is
\begin{eqnarray}
C 
= C_5
= \frac{Z_A \pgamma^A}{l} \pgamma_2 \pgamma_4 \pgamma_5.
\end{eqnarray}
and a dS 'Majorana' spinor 
\begin{eqnarray}
\psi_M 
%= C \bar{\psi}_M^T 
= C_5 \bar{\psi}_M^T.
%= \qgamma_2 \psi_M^*
\label{majorana2}
\end{eqnarray}
We propose a $SO(1,4)$ invariant dS 'Majorana' fermion action
by replacing a Dirac spinor to a dS 'Majorana' spinor
in the action (\ref{dirac2})
\begin{eqnarray}
{\cal L}_{Majorana}
= - \epsilon^{ABCDE} \epsilon^{\mu\nu\rho\sigma}
{\bar \psi_M} 
\left(\frac{Z_A}{l} \pgamma_{B} \frac{\overleftrightarrow{D}_{\mu}}{3!}
+ \lambda \frac{Z_A}{l} \frac{D_{\mu} Z_B}{4!}
\right) \psi_M
D_{\nu} Z_C D_{\rho} Z_D D_{\sigma} Z_E.
\label{majoranaaction2}
\end{eqnarray}
We can prove the consistency of the action (\ref{majoranaaction2}) 
for the charge conjugation $C_5$ similar to $SO(2,3)$ case.

If we break the $SO(1,4)$ symmetry by $Z_A=(0,0,0,0,l)$,
(\ref{majoranaaction2}) becomes
the Majorana fermion action in the Einstein gravitational theory
in four dimensions
\begin{eqnarray}
{\cal L}_{Majorana}
&=& -e {\bar \psi_M} \left(
\pgamma_a
e^{\mu a} \overleftrightarrow{D}_{\mu}
+ \lambda \right) \psi_M,
%\nonumber \\
%&=& -e {\bar \psi_M} \left(
%\pgamma_a
%e^{\mu a} \overleftrightarrow{\circD}_{\mu}
%+ \lambda - \frac{2i}{l} \right) \psi_M
\end{eqnarray}
and a 
%complex 
mass
\begin{eqnarray}
m = \lambda
%- \frac{2i}{l}.
\end{eqnarray}
is obtained in the $SO(1,4)$ case.

%%%%%%%%%%%%%%%%%%%%%%%%%%%%%%%%%%%%%%%%%%%%%%%%%%%%%%%%%%%%%%%%%%%%%
%%%%%%%%%%%%%%%%%%%%%%%%%%%%%%   SEC  8    %%%%%%%%%%%%%%%%%%%%%%%%%%
%%%%%%%%%%%%%%%%%%%%%%%%%%%%%%%%%%%%%%%%%%%%%%%%%%%%%%%%%%%%%%%%%%%%%
\section{Conclusions and Discussions}
\noindent
We have constructed $SO(2,3)$ and $SO(1,4)$ invariant fermion actions
which derive Weyl fermion actions or Majorana fermion actions 
with the AdS and dS 
gravity after breaking the symmetries to $SO(1,3)$.
%after breaking the symmetries to $SO(1,3)$ and 
%deriving the AdS and dS gravities.

The keys ingredients are chiral projection operators and charge conjugations.
In Weyl fermion case, we have constructed
$SO(2,3)$ and $SO(1,4)$ invariant operators $P_{\pm}$ which reduce to 
and chiral projections if we break the symmetries to $SO(1,3)$ in 
Weyl fermions.
In Majorana fermion case, we have constructed that
$SO(2,3)$ and $SO(1,4)$ covariant 'charge conjugations'
$C_2$ and $C_5$ 
which becomes
the $SO(1,3)$ charge conjugation if we break the symmetries
to $SO(1,3)$.

%This new mechanism to derive chiral theories from nonchiral theories
%seems to be useful
%to build new chiral fermions models.
The following remarks are in order.
Our theory arrives finally at a $SO(1,3)$ invariant theory.
However, it is not the same as the theory which is 
$SO(1,3)$ invariant from the start.
For instance, the usual $SO(1,3)$ invariant gravity does not 
allow the metric as the gauge ingredient. The equations
(\ref{23gravity}) and (\ref{14gravity}) are
invariant under $SO(1,3)$ but with the special coefficients
of three terms with topological term, only in which case some 
conserved quantities appear 
\cite{Aros:1999kt}.
The same thing occurs for the case in the presence of the fermions.
This becomes more explicit if we consider the fluctuations 
$Z_A$ from (\ref{breaking1}) and (\ref{breaking2}) 
though it is beyond the scope of this article.

As is easily understood, there are clear differences between the original 
$SO(2,3)$ (or $SO(1,4)$) action and a $SO(1,3)$ action from the 
start even after symmetry breaking of the former. 
The former includes more informations than the latter.
Also equations of motion of the $SO(2,3)$ (or $SO(1,4)$) action 
are different from those obtained by imposing 
(\ref{breaking1}) (or (\ref{breaking2})) before Euler variation
and put the constraints explicitly done in this paper and
\cite{Fukuyama:1982kz}.
Our formulation makes it possible to check the gravitational 
chiral anomaly \cite{Chandia:1997hu}.

The anti de Sitter group and spacetime is important 
also in the context of the AdS/CFT correspondence.
It is interesting 
to analyze relations of the AdS gravity to the AdS/CFT correspondence
\cite{Fukuyama:2009cm}.

%%%%%%%%%%%%%%%%%%%%%%%%%%%%%%%%%%%%%%%%%%%%%%%%%%%%%%%%%%%%%%%%%%%%%
%%%%%%%%%%%%%%%%%%%%%%%%%%%%   ACKNOWLEGE    %%%%%%%%%%%%%%%%%%%%%%%%
%%%%%%%%%%%%%%%%%%%%%%%%%%%%%%%%%%%%%%%%%%%%%%%%%%%%%%%%%%%%%%%%%%%%%
\section*{Acknowledgments}
\noindent
The work of T.F. is supported in part by the grant-in-Aid 
for Scientific Research from the Ministry of Education, 
Science and Culture of Japan (No. 20540282). 
We thank A.Randono for pointing out typos.

%%%%%%%%%%%%%%%%%%%%%%%%%%%%%%%%%%%%%%%%%%%%%%%%%%%%%%%%%%%%%%%%%%%%
\setcounter{section}{0}

%%%%%%%%%%%%%%%%%%%%%%%%%%%%%%%%%%%%%%%%%%%%%%%%%%%%%%%%%%%%%%%%%%%%%
%%%%%%%%%%%%%%%%%%%%%%%%%%%%%%   APPENDIX    %%%%%%%%%%%%%%%%%%%%%%%%%%
%%%%%%%%%%%%%%%%%%%%%%%%%%%%%%%%%%%%%%%%%%%%%%%%%%%%%%%%%%%%%%%%%%%%%

%\section*{Appendix}
\noindent
%In appendix, 
\section*{Appendix: Notation 
%in Four Dimensions
}
\noindent
We summarize formulae of gamma matrices and 
charge conjugations in $SO(1,3)$, $SO(2,3)$ and $SO(1,4)$ groups.
%in our paper.
\setcounter{section}{1}
\subsection{$SO(1,3)$ Metric}
\noindent
We take a local coordinate $(x_1, x_2, x_3, t)$ in four dimensions
and a $SO(1,3)$ invariant metric with the signature $(+, +, +, -)$.
We define an imaginary coordinate $x_4 = it$ and
we use a coordinate $(x_1, x_2, x_3, x_4)$
with the signature $(+, +, +, +)$.

%\hfil\break
\subsection{$SO(2,3)$ Metric}
\noindent
We consider a five vector $(Z_1, Z_2, Z_3, T_1, T_2)$ 
and $SO(2,3)$ is defined as the invariant group
of the norm $Z_1^2 + Z_2^2 + Z_3^2 - T_1^2 - T_2^2$
with the signature $(+, +, +, -, -)$.

We define  $Z_4 = iT_1, Z_5 = iT_2$ and 
we use the notation $(Z_1, Z_2, Z_3, Z_4, Z_5)$
with the signature $(+, +, +, +, +)$.
$Z_5 = i l$ is pure imaginary, 
where $l$ is a real number.

%$\epsilon^{ABCDE} = 1$ 

\subsection{$SO(1,4)$ Metric}
\noindent
We consider a five vector $(Z_1, Z_2, Z_3, T, Z_5)$
and $SO(1,4)$ is defined as the invariant group
of the norm $Z_1^2 + Z_2^2 + Z_3^2 - T^2 + Z_5^2$
with the signature $(+, +, +, -, +)$

We define $Z_4 = iT$ and 
we use the notation $(Z_1, Z_2, Z_3, Z_4, Z_5)$
with the signature $(+, +, +, +, +)$.
$Z_5 = l$ is real, where $l$ is a real number.

%%%%%%%%%%%%%%%%%%%%%%%%%%%%%%%%%%%%%%%%%%%%%%%%%%%%%%%%%%%%%%%%%%%%%%
\subsection{$SO(1,3)$ and $SO(1,4)$ Gamma Matrix}
\noindent
We take gamma matrices $\gamma_a$ in the Lorentz group.
where gamma matrices satisfy
\begin{eqnarray}
\{\pgamma_a, \pgamma_b \} = 2 \delta_{ab},
\end{eqnarray}
where $a=1,2,3,4$.
We call $\gamma_a$ the $SO(1,3)$ gamma matrices.
%In $SO(1,3)$, 
$\pgamma_5$ is defined as
\begin{eqnarray}
\pgamma_5 = - \pgamma_1 \pgamma_2 \pgamma_3 \pgamma_4.
\label{gamma5}
\end{eqnarray}
Then gamma matrices including $\pgamma_5$ satisfy 
\begin{eqnarray}
\{\pgamma_A, \pgamma_B \} = 2 \delta_{AB}.
\label{anticomm}
\end{eqnarray}
where $A = 1, 2, 3, 4, 5$.

$SO(1,4)$ gamma matrices are defined as $\gamma_A = (\gamma_a, \gamma_5)$.

The gamma matrices are hermitian
\begin{eqnarray}
\pgamma_A^{\dag} = \pgamma_A.
\label{hermitian}
\end{eqnarray}
A useful formula is
\begin{eqnarray}
Z_A \pgamma_A Z_B \pgamma_B = 
%\not
\slash{Z} \!\! \slash{Z} = 
Z_A Z_A = Z^2 = l^2.
\end{eqnarray}
The symmetry generators $s_{AB}$ are given by
\begin{eqnarray}
s_{AB} = \frac{1}{4i} [\pgamma_A, \pgamma_B].
\end{eqnarray}
$s_{AB}$ satisfies 
\begin{eqnarray}
s_{AB}^{\dag} = s_{AB}.
\end{eqnarray}
We take the Dirac (Pauli) basis of gamma matrices in this paper,
\begin{eqnarray}
&& \pgamma_1
= i \left(
\begin{array}{cc}
0 & - \sigma_1  \\
\sigma_1 & 0 \\
\end{array}
\right),
\quad
\pgamma_2
= i \left(
\begin{array}{cc}
0 & - \sigma_2 \\
\sigma_2 & 0 \\
\end{array}
\right),
\quad
\pgamma_3
= i \left(
\begin{array}{cc}
0 & - \sigma_3 \\
\sigma_3 & 0 \\
\end{array}
\right),
\nonumber 
\\
&& 
\pgamma_4
= \left(
\begin{array}{cc}
I & 0 \\
0 & -I \\
\end{array}
\right),
\quad
\pgamma_5
= \left(
\begin{array}{cc}
0 & I \\
I & 0 \\
\end{array}
\right),
\end{eqnarray}
where $\sigma_1$, $\sigma_2$ and $\sigma_3$ are the Pauli matrices
and $I$ is the $2 \times 2$ unit matrix.
This basis satisfies
(\ref{anticomm}), (\ref{hermitian})
and 
\begin{eqnarray}
\pgamma_A^T =
\left\{
\begin{array}{cc} 
\pgamma_A & \mbox{if} \qquad A=2, 4, 5, \\ 
- \pgamma_A  & \mbox{if} \qquad A=1, 3,
\end{array}
\right.
\end{eqnarray}
\begin{eqnarray}
\pgamma_A^* =
\left\{
\begin{array}{cc} 
\pgamma_A & \mbox{if} \qquad A=2, 4, 5, \\ 
- \pgamma_A  & \mbox{if} \qquad A=1, 3. 
\end{array}
\right.
\end{eqnarray}
Another basis,
the Weyl (chiral) basis of gamma matrices, is
\begin{eqnarray}
&& \pgamma_1
= i \left(
\begin{array}{cc}
0 & \sigma_1  \\
- \sigma_1 & 0 \\
\end{array}
\right),
\quad
\pgamma_2
= i \left(
\begin{array}{cc}
0 & \sigma_2 \\
- \sigma_2 & 0 \\
\end{array}
\right),
\quad
\pgamma_3
= i \left(
\begin{array}{cc}
0 & \sigma_3 \\
- \sigma_3 & 0 \\
\end{array}
\right),
\nonumber 
\\
&& 
\pgamma_4
= \left(
\begin{array}{cc}
0 & I \\
I & 0 \\
\end{array}
\right),
\quad
\pgamma_5
= \left(
\begin{array}{cc}
I & 0 \\
0 & -I \\
\end{array}
\right).
\end{eqnarray}
This basis is useful to consider a Weyl fermion.

%%%%%%%%%%%%%%%%%%%%%%%%%%%%%%%%%%%%%%%%%%%%%%%%%%%%%%%%%%%%%%%%%%%%%%
%\subsection{Definition of $\bar{\psi}$}
%\noindent
%\subsection{4D $SO(1,3)$ case}
%\noindent
Let $\psi$ be a Dirac spinor for $SO(1,3)$ or $SO(1,4)$.
${\bar \psi}$ is defined as
${\bar \psi} = \psi^{\dag} \pgamma^4$ for both
$SO(1,3)$ and $SO(1,4)$.

%%%%%%%%%%%%%%%%%%%%%%%%%%%%%%%%%%%%%%%%%%%%%%%%%%%%%%%%%%%%%%%%%%%%%
\subsection{$SO(2,3)$ Gamma Matrix}
\noindent
We define $SO(2,3)$ 
%and $SO(1,4)$ 
gamma matrices $\qgamma_A$ in the AdS gravity as follows:
\begin{eqnarray}
&& 
\qgamma_a \equiv - i \pgamma_5 \pgamma_a,
\nonumber \\
&& 
\qgamma_5 \equiv \pgamma_5.
\end{eqnarray}
Then
\begin{eqnarray}
&& 
\pgamma_a \equiv i \qgamma_5 \qgamma_a,
\nonumber \\
&& 
\pgamma_5 \equiv \qgamma_5.
\end{eqnarray}
$\qgamma_A$ satisfies 
\begin{eqnarray}
&& \{\qgamma_A, \qgamma_B \} = 2 \delta_{AB},
\nonumber \\
%\end{eqnarray}
%and
%\begin{eqnarray}
&& \qgamma_A^{\dag} = \qgamma_A.
\end{eqnarray}
We have taken the Dirac basis for $SO(1,3)$ gamma matrices $\pgamma_A$.
Then $\qgamma_A$ satisfy
\begin{eqnarray}
\qgamma_A^T =
\left\{
\begin{array}{cc} 
- \qgamma_A & \mbox{if} \qquad A=2, 4, \\ 
\qgamma_A  & \mbox{if} \qquad A=1, 3, 5,
\end{array}
\right.
\end{eqnarray}
\begin{eqnarray}
\qgamma_A^* =
\left\{
\begin{array}{cc} 
- \qgamma_A & \mbox{if} \qquad A=2, 4, \\ 
\qgamma_A  & \mbox{if} \qquad A=1, 3, 5.
\end{array}
\right.
\end{eqnarray}
The symmetry generators given by
\begin{eqnarray}
S_{AB} = \frac{1}{4i} [\qgamma_A, \qgamma_B],
\end{eqnarray}
 satisfy
the following identities:
\begin{eqnarray}
S_{AB}^T =
\left\{
\begin{array}{cc} 
S_{AB} & \mbox{if} \qquad 
(A, B) = (1, 2), (1, 4), (2, 3), (2, 5), (3, 4), (4, 5),  \\ 
- S_{AB} & \mbox{if} \qquad 
(A, B) = (1, 3), (1, 5), (2, 4), (3, 5), \\ 
\end{array}
\right.
\end{eqnarray}
\begin{eqnarray}
&& S_{AB}^{\dag} = S_{AB}.
\nonumber \\
%\end{eqnarray}
%\begin{eqnarray}
&& S_{ab} = s_{ab}.
\end{eqnarray}
There is a useful formula
\begin{eqnarray}
Z_A \pgamma_A Z_B \pgamma_B 
= \slash{Z} \!\! \slash{Z}
= Z^2 = -l^2.
\end{eqnarray}

%%%%%%%%%%%%%%%%%%%%%%%%%%%%%%%%%%%%%%%%%%%%%%%%%%%%%%%%%%%%%%%%%%%%%%
%\subsection{Definition of $\bar{\psi}$}
%\noindent
%\subsection{4D $SO(1,3)$ case}
%\noindent
Let $\psi$ be a Dirac spinor for the $SO(2,3)$ group.
We define ${\bar \psi}$ as
${\bar \psi} = \psi^{\dag} \qgamma^5 \qgamma^4$.

%\vfil\eject
%%%%%%%%%%%%%%%%%%%%%%%%%%%%%%%%%%%%%%%%%%%%%%%%%%%%%%%%%%%%%%%%%%%%%
%%%%%%%%%%%%%%%%%%%%%%%%%%%%%%%%%%%%%%%%%%%%%%%%%%%%%%%%%%%%%%%%%%%%%
%%%%%%%%%%%%%%%%%%%%%%%%%%%%%%%%%%%%%%%%%%%%%%%%%%%%%%%%%%%%%%%%%%%%%

\newcommand{\bibit}{\sl}
%%%%%%%%%%%%%%%%%%%%%%%%%%%%%%%%%%%%%%%%%%%%%%%%%%%%%%%%%%%%%%%%%%%%%
%%%%%%%%%%%%%%%%%%%%%%%%%%%%%%%  Refs. %%%%%%%%%%%%%%%%%%%%%%%%%%%%%%
%%%%%%%%%%%%%%%%%%%%%%%%%%%%%%%%%%%%%%%%%%%%%%%%%%%%%%%%%%%%%%%%%%%%%
%\newpage
%NEW MACRO FOR BIBLIOGRAPHY

%\section*{References}
%\noindent

\vfill\eject
\end{document}